\documentclass[twocolumn,amsmath,amssymb,prl]{revtex4}
\usepackage{amsmath}
\usepackage{amssymb}
\usepackage[dvips]{graphicx}
\usepackage{color} 
\usepackage{moreverb}
\usepackage{bbm}
\usepackage[plainpages=false,pdftex,colorlinks=false,pdfborder={0 0 0}]{hyperref}
\begin{document}
\title{Spin Injection Enhancement Through Schottky Barrier Superlattice Design}
\author{Joseph Pingenot}
\altaffiliation[Current address: ]{Department of Physics, University of Oklahoma, Norman, OK 73019}
\author{Michael E. Flatt\'e}
\email{michael_flatte@mailaps.org}
\affiliation{Optical Science and Technology Center and Department of Physics and Astronomy, University of Iowa, Iowa City, IA 52242}
\date{\today}
\begin{abstract}
We predict it is possible to achieve high-efficiency room-temperature spin injection from a mag- netic metal into InAs-based semiconductors using an engineered Schottky barrier based on an InAs/AlSb superlattice. The Schottky barrier with most metals is negative for InAs and positive for AlSb. For such metals there exist InAs/AlSb superlattices with a conduction band edge perfectly aligned with the metal's Fermi energy. The initial AlSb layer can be grown to the thickness required to produce a desired interface resistance. We show that the conductivity and spin lifetimes of such superlattices are sufficiently high to permit efficient spin injection from ferromagnetic metals.
\end{abstract}
\maketitle

Spin injection from magnetic metals into semiconductors, an important step for spintronic devices\cite{Awschalom2007,Awschalom2002}, has been demonstrated for both a Schottky barrier and an oxide tunnel barrier\cite{Hanbicki2002,Motsnyi2002,Erve2004,Adelmann2005}.  Measurements of spin injection from iron into GaAs at low temperature, and with an applied growth-direction magnetic field, have approached the expected ideal efficiency\cite{Jonker2006}. Spin injection into InAs has proved more challenging, typically with low efficiency reported even at low temperature\cite{Ohno2003,Hammar2002,Hu2001}, and a rapid reduction in efficiency as the temperature is increased. Yet the InAs/GaSb/AlSb material system offers significant advantages over GaAs-based systems, including very large $g$-factors\cite{Yu3ed} and a broad range of tunability of the spin lifetimes with electric fields\cite{Hall2003}. A challenge for spin injection into InAs is that InAs (unlike GaAs) commonly forms an Ohmic contact with metals.
An Ohmic contact severely limits the spin injection efficiency because of the mismatch between the conductivity of the magnetic metal and the conductivity of the semiconductor\cite{Schmidt2000}. One known solution is to use a tunnel barrier at the interface\cite{Rashba2000} to generate a large spin-dependent resistance. With carefully controlled doping, the naturally-occurring Schottky barriers between  metals and semiconductors will behave as such a tunnel barrier. The larger the doping levels used, the thinner the barrier, and the higher the conductivity of the barrier. High doping levels have an undesirable effect, however, on the spin lifetime of a nonmagnetic semiconductor\cite{Kikkawa1998}, whereas at lower doping levels the spin-polarized carriers accelerate in the remaining built-in field from the Schottky barrier, also enhancing spin relaxation\cite{Sanada2002}.
%The large electric fields near that interface produced by the doping can also have either a positive\cite{Yu,Yu2} or negative\cite{Albrecht} effect on spin injection efficiency.

Here we describe an approach to spin injection which permits very efficient spin injection from a magnetic metal into InAs-based semiconductors. We replace the Ohmic interface between the metal and InAs with an InAs/AlSb superlattice. The layer thicknesses of this superlattice are chosen to match the conduction band energy of the superlattice to the Fermi energy of the metal. Thus no doping is required to reduce the Schottky barrier (as would be required with either GaAs or thick AlSb). Spin lifetimes for a (110)-grown InAs/AlSb superlattice, electron doped to $n=10^{17}$~cm$^{-3}$, exceed 1~ns at room temperature, so spins do not decohere significantly during their transit through the superlattice. The vertical conductivity of this superlattice exceeds 1 $\Omega$ cm. These structures may form the spin-injection contacts for a spin transistor such as described in Ref.~\onlinecite{Hall2003,Hall2006}. To calculate the spin injection quantitatively we assume parameters for the magnetic metal corresponding to cobalt.

\begin{figure}[htp]
\includegraphics[width=\columnwidth]{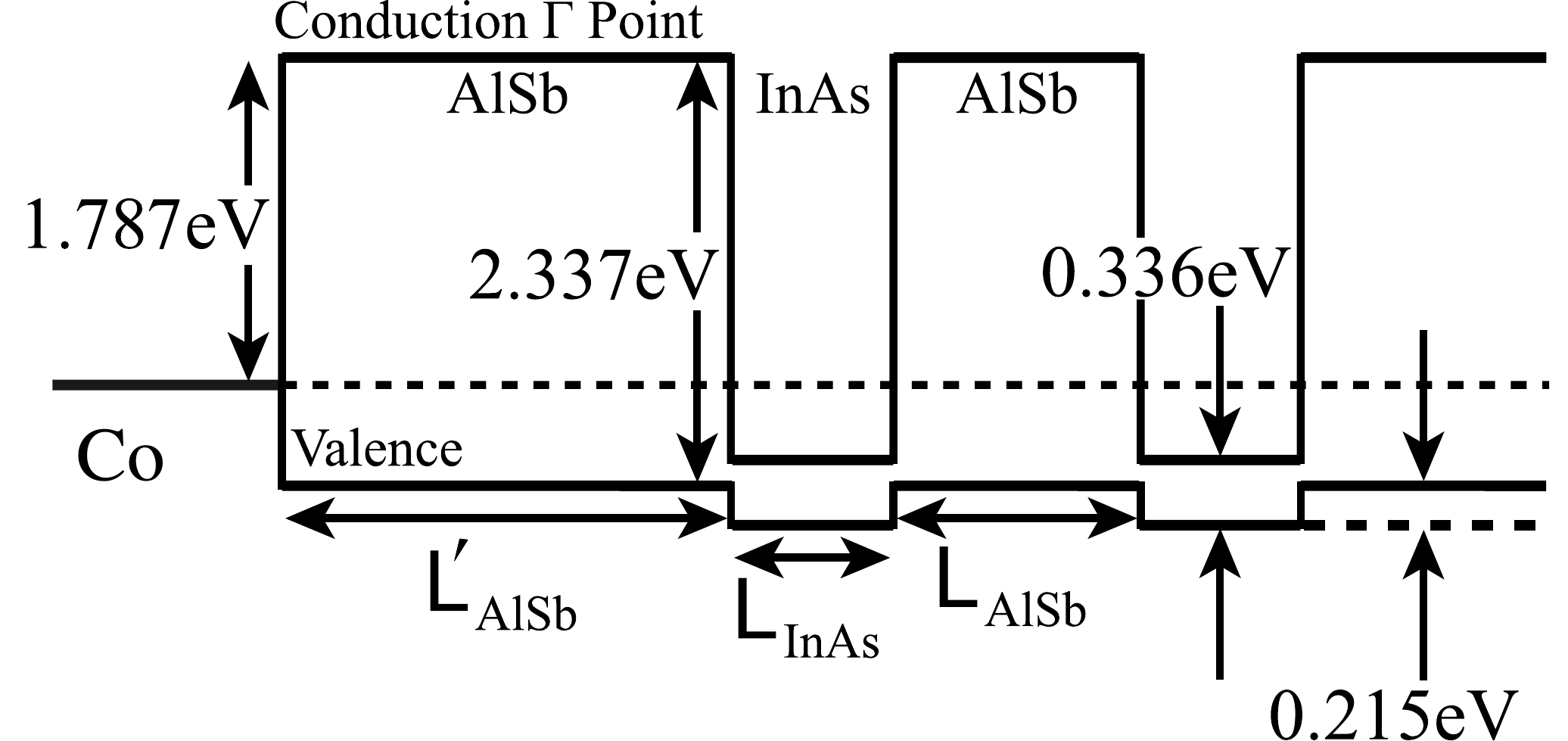}
\caption{Schematic of the band edges of the constituent materials of the superlattice.  $L_{\rm InAs}$ and $L_{\rm AlSb}$ are chosen to match the superlattice conduction band edge (dashed line) to the Co Fermi level, whereas $L'_{\rm AlSb}$ controls the interface resistance.}\label{slillus}
\end{figure}

Schottky barriers between two materials form because the work functions of the isolated materials differ. The resulting charge transfer required to make the work functions of the two materials equal far from the barrier produces the built-in electric fields characteristic of the Schottky barrier. Usually the work function of the semiconductor is smaller than that of the metal, however the Ohmic contact between a metal and $n$-type InAs forms because the work function of $n$-type InAs is greater than that of the metal. The work function of an $n$-type superlattice material with a well (InAs) and barrier (AlSb) is reduced by the confinement energy of the first conduction subband (that is, by the energy of that state above the InAs bulk conduction band edge). Thus if the superlattice well has a work function larger than the metal, and the superlattice barrier has a work function smaller than the metal, then the superlattice work function can be matched to that of the metal by appropriate choice of superlattice layer thicknesses.  No significant charge transfer or built-in electric field occurs when the work functions of the $n$-type superlattice and the metal are matched this way. The interface resistance should be controllable by changing the thickness of the first (superlattice barrier) layer which is in contact with the metal.

The Schottky barrier between a metal and a semiconductor is largely independent of the metal, due to the short screening length inside the metal. For AlSb the typical Schottky barrier is 0.55~eV\cite{Sze1981} ({\it e.g.} as measured for Au/AlSb\cite{Mead1964}), and as illustrated in Figure \ref{slillus}. The valence band offset between InAs and AlSb (Fig.~\ref{slillus}) is $0.215$~eV.  To determine the $n$-type InAs/AlSb superlattices whose work functions match the metal we calculated the band structure of InAs/AlSb superlattices on a grid of layer thicknesses (with 1 monolayer spacing) using a fourteen-band ${\bf k}\cdot {\bf p}$ theory that has proved successful in calculating spin lifetimes, optical matrix elements, conductivities, and carrier lifetimes in a wide variety of superlattices and quantum wells\cite{Lau2004}. From this grid we linearly interpolated AlSb and InAs layer thicknesses precisely corresponding to the correct conduction band edge energy. The the growth direction and temperature can have a profound effect on the superlattice properties, so we repeated the above process for [001] and [110] growth directions, at both 300K and 77K.

\begin{figure}[htp]
\includegraphics[width=\columnwidth]{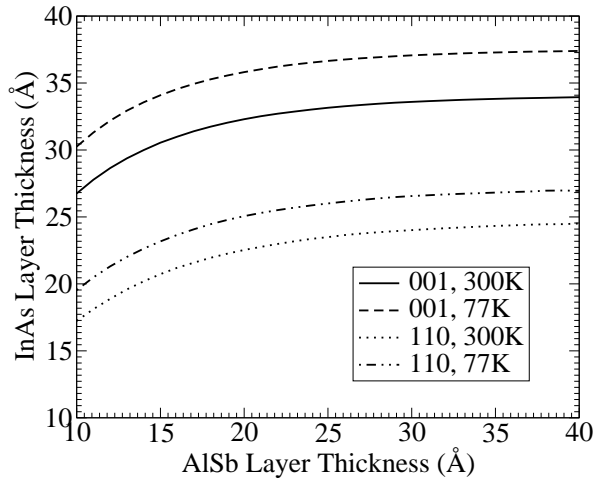}
\caption{Superlattice layer widths corresponding to an $n$-type InAs/AlSb superlattice with a work function matching that of cobalt.}\label{matches}
\end{figure}

Figure \ref{matches} shows the superlattice layer thicknesses whose conduction band lies at the correct energy.  As one would expect, as the AlSb thickness increases, the InAs quantum wells (see figure \ref{slillus}) become more and more isolated between thicker and thicker AlSb barriers, leading to a limiting value of the InAs layer thickness.  Of note is also the small difference (approximately 3-4\AA) between 77K and 300K.  The values shown are similar to those previously found to produce low InAs/AlSb Schottky barriers for low-power electronics\cite{Chow1996}.

\begin{figure}[htp]
\includegraphics[width=\columnwidth]{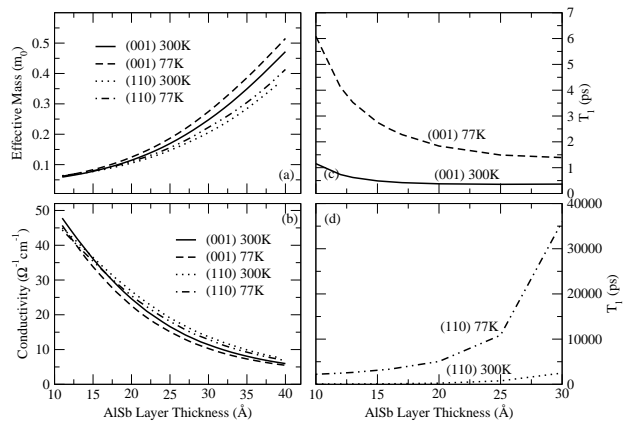}
\caption{Masses[a], Conductivities[c], and spin lifetimes ($T_1$) for (001)-grown[b] and (110)-grown[d] superlattices.}\label{quadgraph}
\end{figure}

It would also be desirable for the superlattice to be conductive and have a long carrier spin lifetime.  Thus we calculated the masses and spin lifetimes $T_1$ for a series of superlattices with layer thicknesses corresponding to the solutions in Fig.~\ref{matches}.  The effective masses came from the fourteen-band  ${\bf k}\cdot{\bf p}$ calculations\cite{Lau2004}, whereas the $T_1$'s were calculated assuming (1) a precessional decoherence (D'yakonov-Perel') mechanism\cite{Dyakonov1972,Lau2004}, and (2)  neutral impurity scattering with an orbital scattering time of $100$~fs dominates the mobility of the superlattice. Figure \ref{quadgraph} shows the masses, calculated conductivities, and $T_1$ times for electrons in (001)- and (110)-grown superlattices corresponding to the solutions in Fig.~\ref{matches}.
  The conductivity depends on the mass according to
\begin{equation}
  \sigma = \frac{ne^{2} \tau_{p} }{m^{*}},\label{mass}
\end{equation}
where $\tau_{p}$, the scattering time from neutral impurities, is assumed (as before in the $T_1$ calculations) to be $100$~fs.
Here $e$ is the charge of the electron
and $n$ is the carrier density (taken to be $10^{17}$~cm$^{-3}$).  From Eq.~(\ref{mass}) the conductivity varies inversely with the electron effective mass, and this is evident in Fig.~\ref{quadgraph}.  

As found generically in bulk and superlattice structures when the spin lifetime is dominated by precessional decoherence, the spin lifetimes of electrons at 77K are much longer than those at 300K.  
The spin lifetimes increase with shorter barrier thicknesses for (001)-grown structures, but they increase with {\it longer} barrier thicknesses for (110)-grown structures.  Also, the spin lifetimes for carriers in the (110)-grown superlattices are orders of magnitude longer than those in the (001)-grown structures, as found in previous calculations\cite{Dyakonov1986}.  Thus spin injection into (110)-grown InAs/AlSb superlattices should be much more efficient than into (001)-grown superlattices, as the spin lifetimes (and hence spin flip lengths) should be longer for (110)-grown superlattices than for (001)-grown superlattices.

Finally, varying the thickness of the first barrier layer ($L'_{\rm AlSb}$) in a superlattice does not affect the band structure of the semi-infinite superlattice.  On the other hand, varying the thickness of this first barrier layer has a substantial effect on the probability of an electron tunneling through the layer, which in turn influences the resistivity of the metal-superlattice interface.  The tunneling resistance of the metal-AlSb interface $r_b$ depends on the thickness $L'_{\rm AlSb}$ according to\cite{Capasso1990}
\begin{equation}
r_b = ({e\pi v_{F}g_m(\epsilon_{F})})^{-1}\exp\left( 2L'_{\rm AlSb}\sqrt{\frac{2m^{*}V_{0}}{\hbar ^2}}\right),\label{tunnel}
\end{equation}
where $V_{0}=0.55$~eV is the AlSb
barrier height, $v_{F}=3.3\times 10^5$~m/s is the Fermi velocity in Co\cite{Petrovykh1998}, and
$g_m(\epsilon_{F})=1.04\times 10^{23}/$eVcm$^{3}$ is the density of states of Co at the Fermi energy \cite{Batallan1975}.

This interface resistance, along with other metal and superlattice parameters, determines the spin polarization of the current through the initial AlSb layer\cite{Fert2001}:
\begin{equation}
\frac{J_\uparrow - J_\downarrow}{J} = \frac{\beta r_{Co} + \gamma r_{b}}{r_{Co} + r_{SL} + r_{b}},\label{initialspin}
\end{equation}
where $\beta$ is the current spin polarization for Co at the Co/AlSb interface, $\gamma$ is the spin polarization of states at the Co interface involved in tunneling through the barrier, and $r_{\rm Co}$ is the resistance of Co over a thickness of the spin difusion length, $r_{\rm Co} = \rho_{\rm Co}l_{\rm Co}^{SF}$, where $\rho_{\rm Co}$ is the resistivity of Co and $l_{\rm Co}^{SF}$ is the spin flip length within Co. 
The superlattice resistance 
$r_{SL} =  (DT_1)^{1/2}/\sigma$,
where $\sigma$ is the conductivity of the superlattice [Eq.~(\ref{mass})] and $D$ is the vertical diffusion
constant of the superlattice,
\begin{equation}
  D = -\frac{\mu}{e}\frac{\displaystyle{\int f(E)g_s(E)dE}}{\displaystyle{\int \frac{\partial f(E)}{\partial E}g_s(E)dE}},
\end{equation}
where $g_s(E)$ is the density of states within the superlattice at energy $E$.

Experimental determinations of $\beta$, $\gamma$, $\rho_{\rm Co}$, and $l_{\rm Co}^{SF}$ differ widely from measurement to measurement. Even a bulk-like quantity like $\beta$ depends on the nonmagnetic material the spin-polarized flows into, and information is not available for a Co/AlSb interface. Thus we use results from the best-characterized metal/metal surface, Co/Cu, for all four of these quantities at 77K and 300K, from a study\cite{Piraux1998} in which all these quantities are simultaneously determined for a set of samples. The values for $\beta$ are 0.36 at 77K and 0.31 at 300K, for $\gamma$ are 0.85 at both temperatures, for $\rho_{\rm Co}$ are $18\mu\Omega$cm at 77K and $25\mu\Omega$cm at 300K, and for $l_{\rm Co}^{SF}$ are $59$~nm at 77K and $38$~nm at 300K.

\begin{figure}[htp]
\includegraphics[width=\columnwidth]{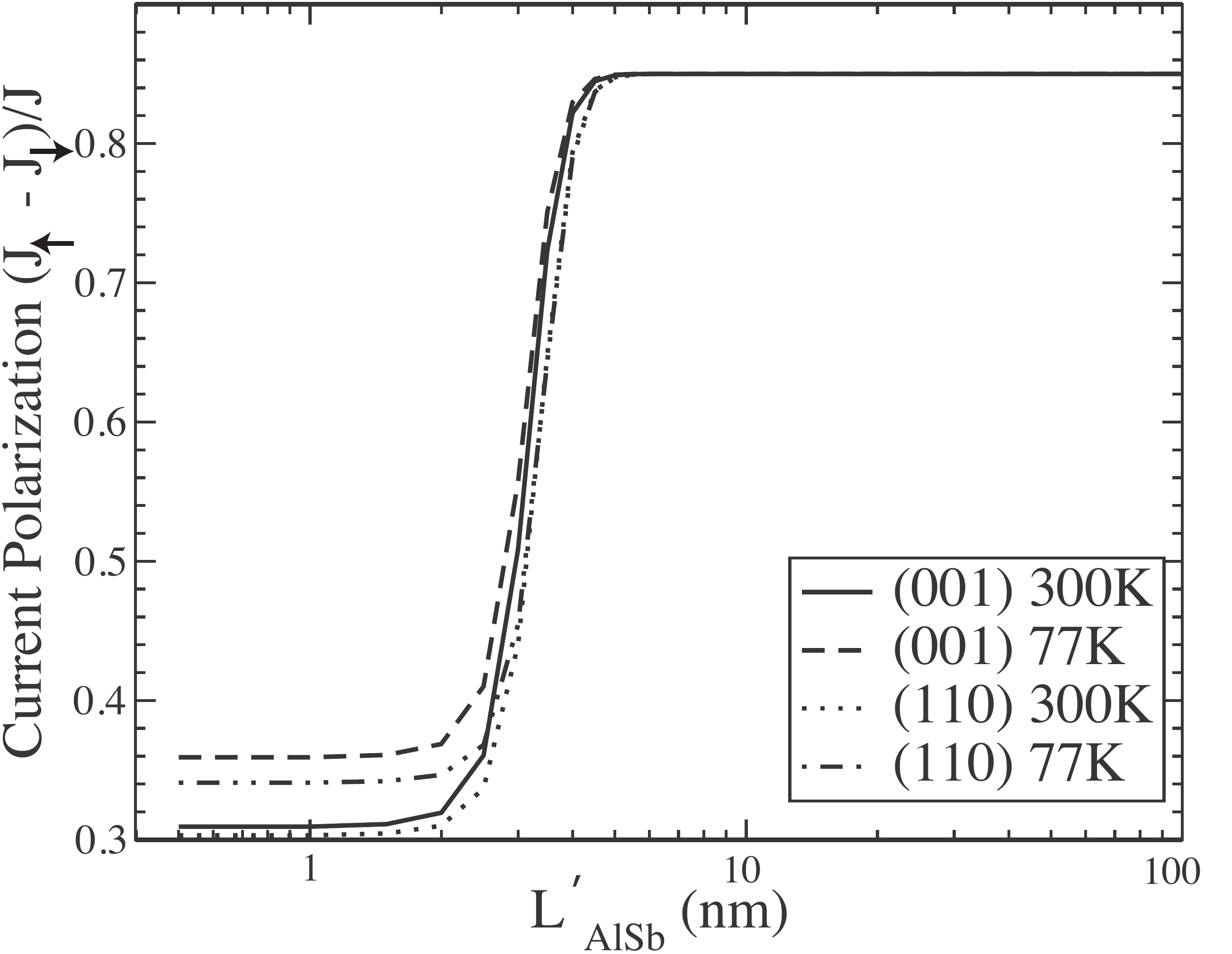}
\caption{Current spin polarization as a function of initial AlSb barrier thickness}\label{si_efficiency}
\end{figure}

Calculations of the initial current spin polarization at the semiconductor interface ($P$) are shown in Fig.~\ref{si_efficiency} for varying initial AlSb barrier thicknesses ($L_{\rm AlSb}'$) but fixed $L_{\rm AlSb}=12$~\AA.  The efficiency approaches the limiting value of $\gamma$ for a thick first layer of AlSb, and the limiting value of $\beta$ for a very thin first layer of AlSb. Thus high-efficiency spin injection is possible, even at room temperature.  As expected from Eq.~(\ref{initialspin}), the initial spin polarization is larger for the (001) superlattice at both temperatures, as the $T_1$ for that superlattice is shorter than that of the (110) superlattice (and thus the $r_{SL}$ for the (001) superlattice is smaller than the $r_{SL}$ for the (110) superlattice). For moderately thick (3~nm) initial AlSb layer thicknesses, however, the initial current spin polarizations for (110) superlattices are as high as can be achieved for these values of $\beta$ and $\gamma$. If the current spin polarizations are comparable for the (001) and (110) superlattices, then spin injection into the (110) superlattices will be more desirable, for the spins will persist in a (110) superlattice for orders of magnitude longer times than in a (001) superlattice (see Fig.~\ref{quadgraph}).

We have described a method of using AlSb/InAs superlattices to enhance spin injection from a magnetic metal such as Co into an InAs-based semiconductor.  The resistivity of the first AlSb barrier can be tuned, while leaving the superlattice's band structure effectively the same.  The matching of the conduction band state of the superlattice to the Fermi level of the metal permits the control of a barrier resistance independently of doping. Current spin injection polarizations of $\sim 0.5$ are predicted to be possible into superlattices with room-temperature $T_1$'s of a few nanoseconds. It should then be possible to grade the superlattice layer thicknesses to match to any desired 6.1\AA\ lattice constant material, including InAs.

\begin{acknowledgments}
This research was supported by the ARO MURI Grant No. W911NF-08-1-0317.  J.P. acknowledges support from the Center for Semiconductor Physics in Nanostructures, an NSF-MRSEC, Grant No. DMR-0520550. We acknowledge numerous helpful discussions with W. H. Lau.
\end{acknowledgments}

\end{document}